\documentclass[twocolumn,preprintnumbers,amsmath,amssymb,superscriptaddress]{revtex4}
\usepackage{epsfig,bm}
\usepackage{ulem}
\usepackage{amsmath,color}

\begin{document}
\title{Utility of antiproton-nucleus scattering
  for probing nuclear surface density distributions}
\author{K. Makiguchi}
\affiliation{Department of Physics,
  Hokkaido University, Sapporo 060-0810, Japan}
\author{W. Horiuchi}
\email{whoriuchi@nucl.sci.hokudai.ac.jp}
\affiliation{Department of Physics,
  Hokkaido University, Sapporo 060-0810, Japan}
\author{A. Kohama}
\affiliation{RIKEN Nishina Center, RIKEN, Wako-shi, Saitama 351-0198, Japan}

\begin{abstract}
Antiproton-nucleon ($\bar{p}N$) total cross sections are typically
3--4 times larger than the $NN$ ones
at incident energies from a few hundreds to thousands MeV.
We investigate antiproton-nucleus scattering as it could work as
a probe of the nuclear structure giving
the sensitivity differently from  a proton probe. 
High-energy antiproton-nucleus reactions 
are reasonably described by the Glauber model
with a minimal profile function that reproduces
the $\bar{p}N$ and $\bar{p}$-$^{12}$C cross section data.
In contrast to the proton-nucleus scattering, 
we find that the complete absorption occurs
even beyond the nuclear radius
due to the large $\bar{p}N$ elementary cross sections,
which shows stronger sensitivity to the nuclear density distribution
in the tail region. This sensitivity is quantified in
the total reaction cross sections with various density profiles
for future measurement including neutron-rich unstable nuclei.
\end{abstract}
\maketitle

\section{Introduction}

Exploring the exotic structure of neutron-rich unstable
  nuclei around the dripline has been one of the main topics
  in nuclear physics.
  Especially, the halo nucleus, which has dilute density
  distributions beyond the nuclear surface,
  appears at around the dripline and has been intensively studied
  since the first discovery of the halo structure
  in $^{11}$Li~\cite{Tanihata85}.
  Probing such density profiles has of particular
  importance to unveil the halo formation mechanism
  as various types of one- and two-neutron halo nuclei have been
  discovered~\cite{Tanihata13}. 
  Recently, a large matter radius of $^{29}$F was observed~\cite{Bagchi20}.
  The structure of the F isotopes at around the dripline
  has attracted attention and already stimulated several theoretical
  works~\cite{Michel20,Masui20,Singh20,Lorenzo20}.

Nuclear density distributions are basic properties of atomic nuclei.
Traditionally, the charge density distributions have been measured 
by using the electron scattering and revealed the nuclear saturation 
properties at internal density distributions~\cite{deVries87}. 
Hadronic probes have also been used to study the nuclear density distributions,
especially at around the nuclear surface. 
Proton-nucleus scattering has been successful 
in determining the matter density distributions of stable nuclei. 
By measuring the elastic scattering differential cross sections 
up to backward angles, detailed nuclear density profiles 
were extracted giving a best fit to the experimental 
cross sections~\cite{Terashima08,Zenihiro10,Sakaguchi17}.

Characteristics of high-energy hadron-nucleus collisions mostly
stem from their elementary processes, more specifically,
hadron-nucleon total cross sections.
For example, proton-neutron ($pn$) and proton-proton ($pp$)
total cross sections have different incident energy dependence, 
especially, at low incident energies~\cite{PDG}.
As was shown in Refs.~\cite{Horiuchi14,Horiuchi16},
this property can be used to extract the proton and neutron radii
as well as the density distributions separately
at around the proton and neutron surfaces~\cite{Hatakeyama18}.
Examining the properties of the other hadronic probes
is interesting as they could be used to extract more information
on the nuclear structure other than the proton probe.
Here we consider high-energy antiproton-nucleus ($\bar{p}A$) scattering.
Note that new experiment
to use the low-energy antiproton beam for studying exotic nuclei
was proposed~\cite{PUMA1, PUMA2}.
At incident energies from 100 MeV to 1 GeV,
elementary cross sections, i.e.,
antinucleon-nucleon ($\bar{N}N$) total cross sections,
are typically 3--4 times larger than those of 
the nucleon-nucleon ($NN$) total cross sections~\cite{PDG}.
With such large cross sections, the $\bar{p}A$ reaction
becomes more absorptive than that of the $pA$
one~\cite{Lichtenstadt85,Kohama16}.
Though the information about the internal
region of the target nucleus is masked by the strong
absorption~\cite{Friedman86},
the antiproton would give different sensitivity to
the nuclear density distributions in the outer regions
compared to that of the proton.

In this paper, we study the high-energy $\bar{p}A$ scattering
to explore the possibility of being a probe of the nuclear structure,
especially focusing on the nuclear
  surface density distributions towards applications
  for studying the exotic structure of neutron-rich unstable nuclei.
The total reaction and elastic scattering cross sections
involving an antiproton as well as a proton
are calculated by a high-energy microscopic reaction theory,
the Glauber model~\cite{Glauber}, which is explained in the following section.
The inputs to the theory is the density distribution of a target nucleus and
the profile function that represents the properties
of the $\bar{N}N$ collision.
Section~\ref{profile.sec} describes how we determine 
the profile function for the $\bar{N}N$ scattering
using the available experimental data.
The parameters of the profile function
are determined following
the available $\bar{N}N$ total cross sections
and $\bar{p}$-$^{12}$C total reaction cross section data.
The validity of this parametrization
is confirmed in comparison with the experimental
elastic scattering differential cross section data for known nuclei.
Section~\ref{discussion.sec} discusses
the properties of the antiproton scattering
in detail comparison to the proton one.
What density profiles are actually probed in
the $\bar{p}A$ scattering is quantified
by examining the total reaction cross sections
with various density profiles.
Conclusions are given in Sec.~\ref{conclusion.sec}.

\section{Glauber model for antinucleon-nucleus scattering}
\label{model.sec}

Here we briefly explain the Glauber model~\cite{Glauber},
which successfully describes high-energy nuclear reactions.
In the Glauber model, the evaluation of the optical-phase-shift
function $e^{i\chi}$ is essential.
The total reaction cross section $\sigma_R$ is calculated by
integrating the reaction probability 
\begin{align}
P(\bm{b})=1-|e^{i\chi(\bm{b})}|^2
\label{prob.eq}
\end{align}
over the impact parameter vector $\bm{b}$ as
\begin{align}
\sigma_R&=\int P(\bm{b})\,d\bm{b}.
\end{align}
Also, the elastic scattering differential cross section is 
calculated by 
\begin{align}
\frac{d\sigma}{d\Omega}&=|F(\theta)|^2
\label{cs.eq}
\end{align}
with the elastic scattering amplitude including
the elastic Coulomb term~\cite{Suzuki03}
\begin{align}
  F(\theta)&=F_C(\eta,\theta)\notag\\
  &+\frac{iK}{2\pi}\int e^{-2iKb\sin\frac{\theta}{2}-2i\eta\ln (Kb)}
  \left(1-e^{i\chi(\bm{b})}\right)\,d\bm{b},
\label{samp.eq}
\end{align}
where $K$ is the wave number in the relativistic kinematics,
and $F_C$ denotes the Rutherford scattering amplitude
with the Sommerfeld parameter $\eta$.

The optical phase-shift function, $e^{i\chi(\bm{b})}$,
which appears in Eqs.~(\ref{cs.eq}) and (\ref{samp.eq}),
includes all information on the high-energy hadron-nucleus scattering
within the Glauber model. However, its evaluation is in general demanding
due to the multiple integration
in the Glauber amplitude~\cite{Glauber}. Though direct integration 
methods were developed using a Monte Carlo integration~\cite{Varga02,Nagahisa18}
and a factorization procedure by assuming a Slater-determinant type
wave function~\cite{Bassel68, Ibrahim09,Hatakeyama14,Hatakeyama15},
in this paper, for the sake of simplicity,
we employ the optical-limit approximation (OLA),
which only takes the leading order term
of the cumulant expansion~\cite{Glauber,Suzuki03}
\begin{align}
  i\chi(\bm{b})
  \simeq -\int \rho_N(\bm{r})\Gamma_{\bar{N}N}(\bm{b}-\bm{s})\,d\bm{r},
\label{OLA.eq}
\end{align}
where $\bm{r}=(\bm{s},z)$ with $\bm{s}$
being a two-dimensional vector perpendicular to the beam direction $z$,
$\rho_{N}$ is the nucleon density distribution, 
and $\Gamma_{\bar{N}N}$ is the $\bar{N}N$
profile function which is responsible for describing the $\bar{N}N$ collision.
One can evaluate the $NA$ scattering by replacing
$\Gamma_{\bar{N}N}$ with $\Gamma_{NN}$ whose standard parameter
sets are tabulated in Refs.~\cite{Horiuchi07,Ibrahim08}.
The choice of the $\bar{N}N$ profile function will 
be discussed in Sec.~\ref{profile.sec}.
We note that the OLA works well in many cases of $pA$ scattering
where the higher order terms are negligible~\cite{Varga02, Ibrahim09, Hatakeyama14, Hatakeyama15, Nagahisa18}.

\section{Determination of the profile function}
\label{profile.sec}

\subsection{Antinucleon-nucleon profile function}

\begin{figure}[ht]
\begin{center}
    \epsfig{file=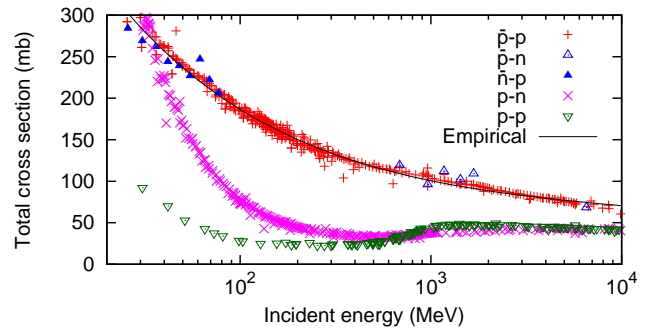, scale=1.1}
    \caption{Antinucleon-nucleon ($\bar{N}N$)
      total cross sections at incident energies from
      30 MeV to 10 GeV. Nucleon-nucleon ($NN$)
total cross sections are also plotted for comparison. 
A curve denotes the empirical parametrization of Eq.~(\ref{totcs.eq}) for 
the $\bar{N}N$ total cross sections used in this paper.}
 \label{NbarN.fig}
\end{center}
\end{figure}

For describing the $\bar{N}A$ scattering,
it is essential to use a reasonable $\Gamma_{\bar{N}N}$
that describes $\bar{N}N$ elementary processes.
Here we take a phenomenological approach to determine 
the $\bar{N}N$ profile function in order to
obtain a global description of the $\bar{N}A$ scattering
in a wide range of the incident energies from few hundred MeV to GeV.
Though it is beyond the scope of this paper,
the construction of the $\bar{N}N$ 
profile function based on the genuine $\bar{N}{N}$ interaction 
is interesting. We note that the recent work~\cite{Vorabbi19}
reproduced the experimental data of $\bar{p}A$ scattering
at about 200 MeV
by using the microscopic wave functions and the $t$-matrix derived
from the $\bar{N}N$ interaction based on the chiral effective field theory.

Here we take the $\bar{N}N$ profile function
as usual finite-range within a Gaussian form~\cite{Ray79}
\begin{align}
  \Gamma_{\bar{N}N}(\bm{b})=\frac{1-i\alpha_{\bar{N}N}}{4\pi\beta_{\bar{N}N}}
  \sigma_{\bar{N}N}^{\rm tot}
  \exp\left(-\frac{\bm{b}^2}{2\beta_{\bar{N}N}}\right),
\label{profile.eq}
\end{align}
where $\sigma_{\bar{N}N}^{\rm tot}$
is the $\bar{N}N$ total cross section,
$\alpha_{\bar{N}N}$ is the ratio of the real to imaginary parts
of the $\bar{N}N$ scattering amplitude at the zero degree,
and $\beta_{\bar{N}N}$ is the slope parameter,
which is responsible for describing
the $\bar{N}N$ elastic scattering at forward angles.
As we will explain it later, these parameters will be fixed to follow
the available $\bar{N}N$ and $\bar{N}A$ scattering data
for each incident energy, although they are limited.

Figure~\ref{NbarN.fig} displays the experimental
$\sigma_{\bar{N}N}^{\rm tot}$ as a function of the incident energy~\cite{PDG}.
The $NN$ total cross sections, $\sigma_{NN}^{\rm tot}$,
are also presented for comparison.
As we see in the figure, $\sigma_{\bar{N}N}^{\rm tot}$ is
approximately 4 times larger than 
$\sigma_{NN}^{\rm tot}$ at around 100 MeV,
and approximately 3 times at around 1000 MeV. These properties
must give the different sensitivity in the $\bar{N}A$ scattering
to the nuclear density profile from that in the $NA$ scattering.
For a practical use, 
we parametrize $\sigma_{\bar{N}N}^{\rm tot}$ in unit of mb
as a function of the incident energy $E$ in unit of MeV
with the same form of Ref.~\cite{Bertulani10} as
\begin{align}
\sigma_{\bar{N}N}^{\rm tot}(E)&=
60.092-361.807/E\notag\\
&+1301.09/\sqrt{E}-2.5882\times 10^{-4}E\notag\\
&(30\,{\rm MeV} <E < 10\,{\rm GeV}).
\label{totcs.eq}
\end{align}
As shown in Fig.~\ref{NbarN.fig}, this empirical parametrization
nicely follows the experimental $\bar{N}N$ data from 30 MeV to 10 GeV.
Since the experimental data are limited, 
especially $\bar{p}n$ scattering cross section data~\cite{PDG}, 
we assume the same value for both $\bar{p}n$
and $\bar{p}p$ cross sections and $\alpha_{\bar{N}N}=0$. 
Note that within the OLA the reaction probability of Eq. (\ref{prob.eq}),
which is the integrand of the total reaction cross section,
does not depend on $\alpha_{\bar{N}N}$ as
$|e^{i\chi(\bm{b})}|^2=e^{-2{\rm Im}\chi(\bm{b})}$.
The remaining parameter
$\beta_{\bar{N}N}$, which determines the effective range of the interaction,
will be fixed in the next subsection.

\subsection{Antiproton-nucleus scattering}

\begin{figure}[ht]
\begin{center}
    \epsfig{file=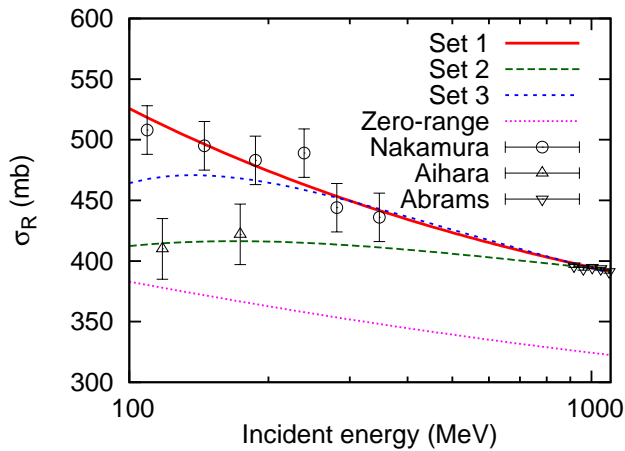, scale=1.4}
    \caption{Total reaction cross sections of $\bar{p}$-$^{12}$C
scattering as a function of incident energy. 
The experimental data are
taken from Refs.~\cite{Abrams71, Aihara81, Nakamura84}.
The density distribution of $^{12}$C is taken
as the harmonic-oscillator-type density distribution~\cite{Ibrahim09}
whose width parameter is fixed so as to to reproduce
the rms point-proton radius extracted from 
the electron scattering~\cite{Angeli13}.}
 \label{pbarC.fig}
\end{center}
\end{figure}

Here we make use of the $\bar{p}$-$^{12}$C total reaction cross section data
to fix $\beta_{\bar{N}N}$ in Eq.~(\ref{profile.eq})
because the density profile of $^{12}$C is well known 
and the elastic scattering differential cross section data
are also available.
By minimizing the root-mean-square (rms)
deviation between the theoretical and experimental
$\bar{p}$-$^{12}$C total reaction cross sections
at different incident energies,
we determine the energy-dependent $\beta_{\bar{N}N}$ parameters.
To obtain a better fit for the experimental cross sections,
we assume $\beta_{\bar{N}N}$ as it similar to the energy dependence
of the total cross sections (\ref{totcs.eq})
\begin{align}
  \beta_{\bar{N}N}(E)=b_1+b_2/E+b_3/\sqrt{E}.
  \label{slope.eq}
\end{align}
Since the experimental data of the 
total reaction cross sections are somewhat scattered at around 200 MeV,
we test three sets of parameters to give the smallest rms
deviation for only with the data of Ref.~\cite{Nakamura84} (Set 1),
of Ref.~\cite{Aihara81} (Set 2), 
and including both cross sections at around 200 MeV (Set 3).
All the potential sets include the data of Ref.~\cite{Abrams71}.

Figure~\ref{pbarC.fig} plots
the calculated total reaction cross sections of the $\bar{p}$-$^{12}$C
scattering with Sets 1--3 as a function of the incident energies.
The results with Sets 1 and 2 are similar
at the incident energies beyond $\approx 200$ MeV,
and the ones with Set 3 show quite differently
from these Sets below $\approx 800$ MeV.
Note that the zero-range profile function
  ($\beta_{\bar{N}N}=0$) defined explicitly by
\begin{align}
  \Gamma_{\bar{N}N}(\bm{b})=\frac{1-i\alpha_{\bar{N}N}}{2}
  \sigma_{\bar{N}N}^{\rm tot}\delta(\bm{b})
\label{profile-zero.eq}
\end{align}
does not explain the experimental data at all, 
giving significant underestimation of the data.
The resulting parameter sets of Eq.~(\ref{slope.eq})
are given in Table~\ref{betapara.tab}.
In general, larger $\beta_{\bar{N}N}$ values $\approx 0.5$--1.3 fm$^2$ are needed
to explain the experimental $\bar{p}$-$^{12}$C data,
while those of the $NN$ profile functions
are ranging from 0.1 to 0.7 fm$^2$~\cite{Horiuchi07}.

\begin{figure}[ht]
\begin{center}
    \epsfig{file=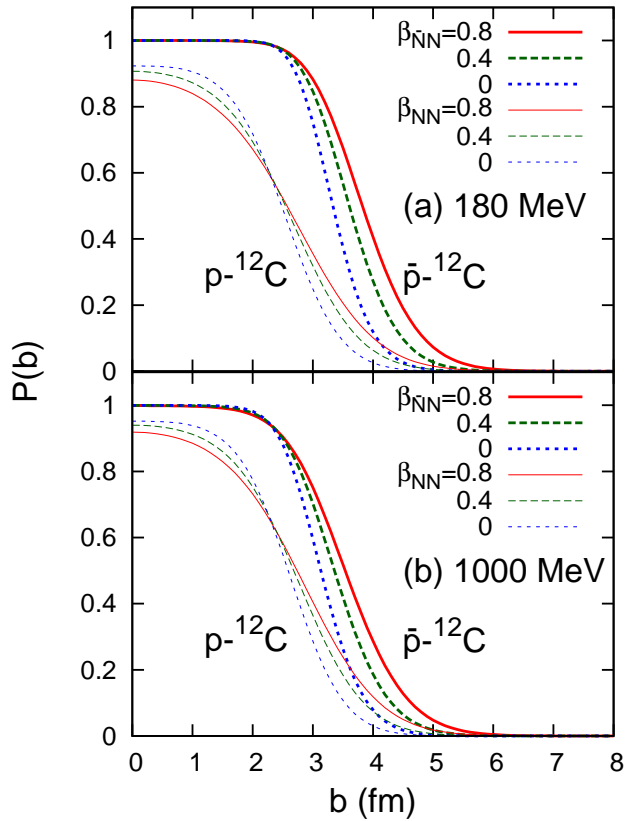, scale=1.4}
    \caption{Reaction probabilities of $\bar{p}$- and $p$-$^{12}$C
in thick and thin lines, respectively, 
as a function of impact parameter $b$
with different slope parameters for the profile functions
at (a) 180 and (b) 1000 MeV.}
 \label{prob.fig}
\end{center}
\end{figure}

These large $\beta_{\bar{N}N}$ values required
in describing the $\bar{N}A$ scattering
can be clarified by comparing the slope parameter 
dependence of the reaction probabilities of Eq.~(\ref{prob.eq})
for the $\bar{p}A$ and $pA$ scattering.
Figure~\ref{prob.fig} draws these reaction probabilities
as a function of the impact parameter $b$ for different 
slope parameters at 180 and 1000 MeV, where the experimental
total reaction cross sections are available.
To compare the role of the slope parameter,
we also take the $NN$ profile function
with $\alpha_{NN}=0$ and vary $\beta_{NN}$.
In the $p$-$^{12}$C scattering,
since the $NN$ total cross section is not large enough
in such a light nucleus, 
the reaction probabilities do not reach 
at unity even at the center of the nucleus ($b=0$), leading to
some slope parameter dependence in the whole regions.
In contrast, in the $\bar{p}$-$^{12}$C scattering,
the probabilities are unity up to around the nuclear radius $\approx 3$ fm.
The tail part of the density distribution beyond the nuclear radius
crucially contributes to the total reaction cross sections.
In fact, the total reaction cross section at 180 MeV 
increases 366, 430, and 487
mb with $\beta_{\bar{N}N}=0.0,$ 0.4, and 0.8 fm$^{-2}$, respectively,
 whereas for the $N$-$^{12}$C scattering, the enhancement is
not as significant as that for the antiproton: 206, 225, and 246
for $\beta_{NN}=0.0$, 0.4, and 0.8 fm$^{-2}$, respectively.
The reaction probabilities at 1000 MeV behave almost
the same as these at 180 MeV with less extended distributions 
because of smaller $\bar{N}N$ total cross sections compared to these at 180 MeV.
Introducing the finite range in the profile function
is essential to describe the $\bar{p}A$ total reaction cross sections.

\begin{table}[ht]
\begin{center}
\caption{Parameters in the parametrization of 
the slope parameter $\beta_{\bar{N}N}$ of Eq.~(\ref{slope.eq}).}
\begin{tabular}{ccccc}
\hline\hline
&&$b_1$ (fm$^2$)&$b_2$ (fm$^2$MeV)&$b_3$ (fm$^2$MeV$^{1/2}$)\\
\hline
Set 1 &&0.4202 &$-4.516\times 10^{-5}$ &4.516\\
Set 2 &&0.7402 & $5.799\times 10^{-5}$&$-$5.799\\
Set 3 &&0.1483 &$-$142.1&17.40\\
\hline\hline
\end{tabular}
\label{betapara.tab}
\end{center}
\end{table}

\begin{figure}[ht]
\begin{center}
    \epsfig{file=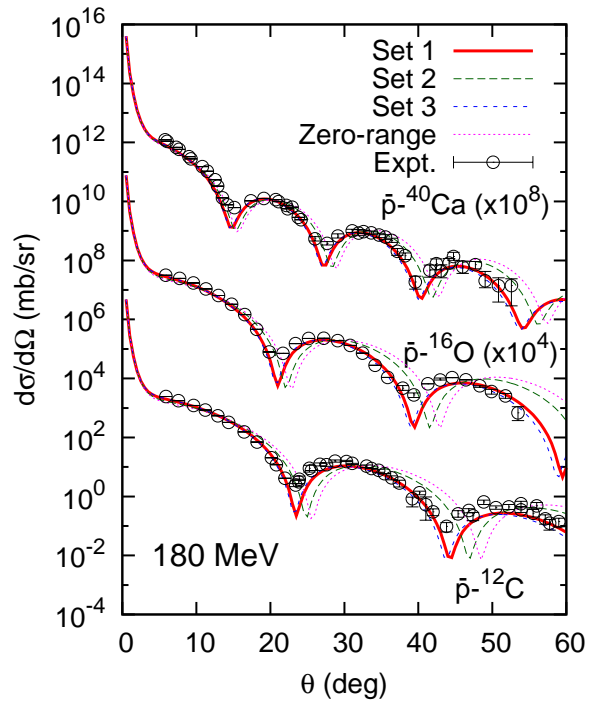, scale=1.5}
    \caption{Elastic scattering differential cross sections
of $\bar{p}$-$^{12}$C, $^{16}$O, and $^{40}$Ca scattering at 180 MeV
with different choices of the profile function.
The experimental data are taken from Ref.~\cite{Garreta84,Bruge86}.
The cross sections are multiplied
by $10^4$ and $10^{8}$ for those for $^{16}$O, and $^{40}$Ca, respectively.
 The harmonic-oscillator-type densities~\cite{Ibrahim09}
 consistent with the experimental charge radii~\cite{Angeli13}
 are used.}
    \label{dcs.fig}
\end{center}
\end{figure}

At the end of this section, the validity of the profile function is examined
by comparing the theoretical elastic scattering
differential cross sections to the experimental data
for $^{12}$C, $^{16}$O, and $^{40}$Ca.
Figure~\ref{dcs.fig} shows the elastic 
scattering differential cross sections
for those target nuclei at the incident energy of 180 MeV.
We find that Set 1 best reproduces the $\bar{p}A$
elastic scattering differential cross section data up to the second minima.
Note that Set 2 also give a good description,
in which its slope parameter
is accidentally almost the same as these of Set 1
at this incident energy region, resulting in the
similar total reaction cross sections shown in Fig.~\ref{pbarC.fig}.
Therefore, we propose the parametrizations of Sets 1 and 2
as a ``minimal'' profile function to describe the $\bar{p}A$ scattering,
and hereafter we use Set 1 otherwise noted.
  While we see overall agreement of the theoretical cross sections with
  the experimental data, at a closer look,
  the cross sections at around the minima are not reproduced well.
  This can be improved by including higher order terms which are ignored
  in the OLA~(\ref{OLA.eq}). See, for example,
  Fig. 1 (a) of Ref.~\cite{Hatakeyama19} for $p$-$^{12}$C scattering.

\section{Discussions}
\label{discussion.sec}

We have confirmed that the $\bar{p}A$
reactions are fairly well reproduced
by the present reaction model.
Here we discuss what density regions are actually probed by the antiproton.
To quantify this, we display the reaction probabilities
of Eq.~(\ref{prob.eq}) as a function of the densities in place of $b$.

\begin{figure}[ht]
\begin{center}
\epsfig{file=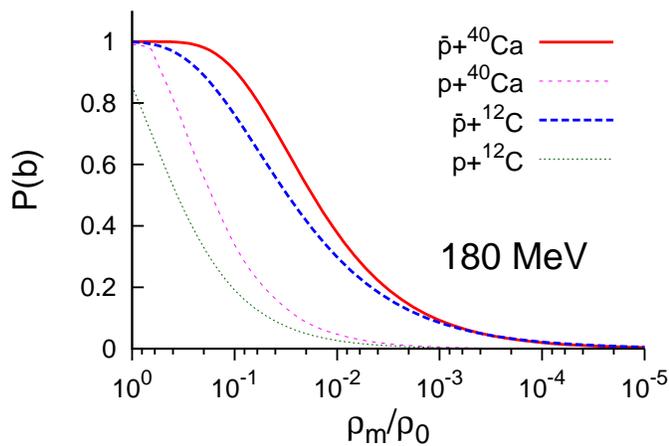, scale=1.4}
\caption{Reaction probabilities of $^{12}$C and $^{40}$Ca
  as a function of the
  fraction of the nuclear density to the central density.
  See text for details.}
 \label{rho-prob.fig}
\end{center}
\end{figure}

Figure~\ref{rho-prob.fig} plots the reaction probabilities of
the $\bar{p}A$ and $pA$ scattering
for $^{12}$C and $^{40}$Ca  at 180 MeV as a function of
the values of $\rho_{m}/\rho_0$, which is the fraction of
the matter density distributions ($\rho_{m}=\rho_n+\rho_p$) to
the density at the origin or the central nuclear density ($\rho_0$).
FOr $^{40}$Ca, at the high density or internal regions,
the probabilities are unity showing the complete absorption
and drop at certain density regions depending on the incident particles.
For the antiproton scattering, the plateau extends
being still unity even at the radius that
the central density is halved $\rho_m/\rho_0$=0.5,
and reaches beyond $\rho_m/\rho_0\lesssim 10^{-4}$,
which is two order of magnitude smaller than that of the proton scattering.
When the probability becomes 0.5,
which corresponds to 5.5 fm of the radius of a sphere,
$\rho_{m}/\rho_0$ becomes 0.02.
This value is one order of magnitude smaller
than that of the proton scattering,
$\rho_{m}/\rho_0=0.2$, corresponding 4.2 fm of the radius of a sphere.
This confirms that the antiproton
can probe the variation of the density distribution at around $\sim 1/100$
of the central density and could be sensitive to the region of
$\rho_{m}/\rho_0\sim 10^{-4}$. This density region corresponds
to the tail of a typical two-neutron halo nucleus
$\rho_{m}/\rho_0\lesssim 10^{-2}$~\cite{Horiuchi06}.

The similar behavior is also found in a case of $^{12}$C.
The plateau also appears for the $\bar{p}$-$^{12}$C,
while the reaction probability
does not reach unity for the $p$-$^{12}$C scattering.
Because the $^{12}$C consists mostly by the nuclear surface,
the optical depth is not small enough.

\begin{figure}[ht]
\begin{center}
\epsfig{file=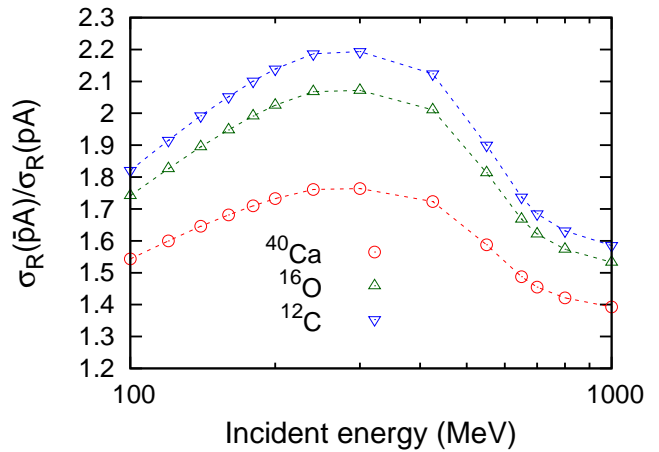, scale=1.4}
\caption{Ratio of total reaction cross sections of $^{12}$C, $^{16}$O, and $^{40}$Ca
  for antiproton and proton scattering as a function of the incident energy.}
 \label{COCa.fig}
\end{center}
\end{figure}

In order to compare the different sensitivity
in the $\bar{p}A$ and $pA$ scattering,
we introduce the ratio of the total reaction cross sections
of the $\bar{p}A$ and $pA$ scattering, $\sigma_R(\bar{p}A) / \sigma_R(pA)$.
The parameter sets of the profile function of Ref.~\cite{Ibrahim08}
are used to calculate $\sigma_R(pA)$.
Let us first discuss a medium-heavy nucleus 
by taking $^{40}$Ca as an example, where the separation of the bulk and 
the surface part is developing~\cite{Kohama05}. 
The curve is shown in Fig.~\ref{COCa.fig}. 
Reflecting the above fact, the antiproton interacts with less nucleons 
than the whole numbers of this nucleus, because the reaction probability 
saturate at the thick density region, i.e., the bulk region, 
as can be seen in Fig.~\ref{rho-prob.fig}.  
The antiproton interacts only with the nucleons in the nuclear surface. 
This is the reason why the ratio does not become large despite the fact
that the $\bar{N}N$ cross section is 3--4 times larger than the $NN$ one. 
What will happen for the case of light nuclei, 
such as $^{12}$C, and $^{16}$O, 
where the nuclear surface is a whole body~\cite{Kohama05}.
Since the most of the composite nucleons are sitting in the surface region, 
the incident antiproton can interact with those nucleons, 
which drastically increases the total reaction cross sections of the antiproton
than that of $^{40}$Ca as one can see from Fig.~\ref{COCa.fig}. 
The energy dependence of the ratio can easily understood by looking
at the values of the elementary cross sections shown in Fig.~\ref{pbarC.fig}.
For example, the $NN$ total cross sections are minimum at this energy region 
while the $\bar{N}N$ ones decreases monotonically,
leading to the peak of the ratio at around 300 MeV.

\begin{figure}[ht]
\begin{center}
\epsfig{file=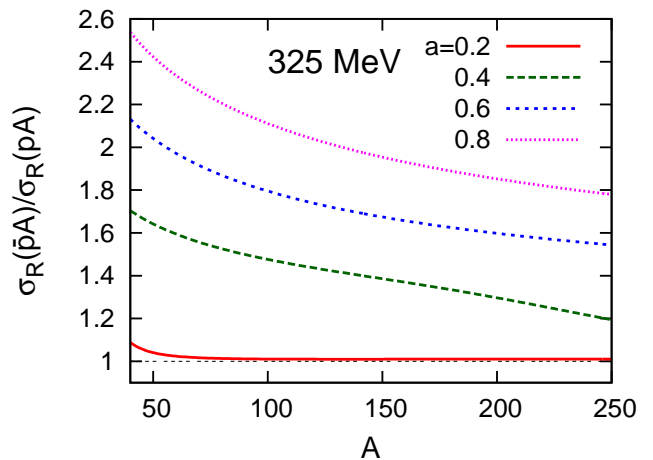, scale=1.4}
\caption{Ratio of total reaction cross sections for antiproton and proton scattering
  at 325 MeV calculated using the 2pF density distributions with various diffuseness parameters
  as a function of the mass number $A$. See text for details.}
 \label{2pFratio.fig}
\end{center}
\end{figure}

To extend the above discussion more general,
we employ two-parameter Fermi (2pF) distributions as a nuclear matter density
\begin{align}
  \rho_N(r)=\frac{\rho_0}{1+\exp\left(\frac{r-R}{a}\right)}.
\label{Fermi.eq}
\end{align}
For a given diffuseness parameter $a$, $\rho_0$ and $R$ are determined
by the normalization to the mass number $A$ and the rms matter radius being
followed as $\sqrt{\frac{3}{5}}1.2\times A^{1/3}$~\cite{Hatakeyama18}.
By varying the diffuseness parameter, we discuss
the role of the surface density profiles for medium to heavy nuclei.
Figure~\ref{2pFratio.fig} plots the calculated cross section ratios with various
diffuseness parameters at 325 MeV, where the ratio is maximized.
Here the averaged $NN$ profile function~\cite{Horiuchi07} is used
to calculate the $NA$ total reaction cross sections.
For a small diffuseness parameter, for example, $a=0.2$ fm,
the ratio is almost unity. Because the reaction probabilities
in the internal regions are already saturated and
a few nucleon exists at around the nuclear surface,
there is no space to increase the total reaction cross sections
even with the larger total $\bar{N}N$ cross sections.
As expected, the smaller diffuseness, the smaller ratio becomes.
We find that the ratio strongly depends on the diffuseness parameter
sufficient to determine the nuclear surface ``diffuseness'' by measuring
both the total reaction cross sections for the $\bar{p}A$ and $pA$ scattering
at the same incident energy.
We note that the typical diffuseness parameters are around 0.45--0.55 fm,
and possibly $\gtrsim 0.6$ fm for well-deformed
and weakly bound nuclei~\cite{Hatakeyama18, Horiuchi17}.
The ratio decreases with increasing the mass number because
the nuclear surface contribution becomes relatively smaller
than the bulk contribution.

\begin{figure}[ht]
\begin{center}
\epsfig{file=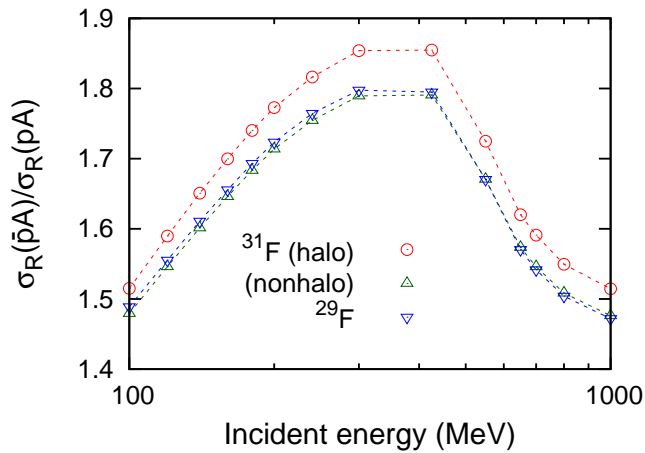, scale=1.4}
\caption{Same as Fig.~\ref{COCa.fig} but of $^{29,31}$F.
  The halo and nonhalo density profiles for $^{31}$F are employed.
  See text for details.}
 \label{2931F.fig}
\end{center}
\end{figure}

Finally, we investigate the different sensitivity 
to a dilute nuclear density profile beyond the nuclear half density radius.
Though the antiproton scattering on unstable nuclei
is still not feasible at present, we take an example of
a possible two-neutron halo nucleus $^{31}$F~\cite{Michel20,Masui20},
which is located at the fluorine dripline~\cite{Ahn19}.
The halo formation depends on the shell gap between
and $0f_{7/2}$ and $1p_{3/2}$ orbits. The inverted configuration,
that is, the dominance of the latter orbit forms the halo structure.
We use these density distributions
of $^{31}$F with the $(1p_{3/2})^2$ (halo) and $(0f_{7/2})^2$ (nonhalo)
dominance, which correspond to the cases A and B in Ref.~\cite{Masui20},
and calculate $\sigma_R(\bar{p}A)/\sigma_R(pA)$
to see the sensitivity to the halo tail.
Figure~\ref{2931F.fig} plots the ratios of $^{31}$F as a function of the incident energy.
For the sake of comparison, the ones of $^{29}$F are also calculated with
the harmonic-oscillator type density distribution~\cite{Masui20}.
The $^{31}$F with the halo tail gives the largest
ratios, while the nonhalo density profile produces the almost
the same behavior of that of $^{29}$F, exhibiting the standard
ratio as expected from Fig.~\ref{COCa.fig}.
This fact clearly shows that the advantage of
the $\bar{p}A$ scattering for the dilute density distribution
further than the nuclear surface.

Since the antiproton has different sensitivity to
the nuclear density profile, one can scan the density distribution by
measuring the elastic scattering differential cross sections
using different probes, the antiproton and proton.
As expected from the diffraction model~\cite{Bethe}
and the recent Glauber model analysis~\cite{Hatakeyama18},
when one performs the antiproton elastic scattering measurement,
elastically scattered particles come to 
the forward angles more concentrated than in the proton case,
which makes the measurement easier.
A detailed study along this direction will give more precise determination
of the nuclear density distributions beyond the nuclear half density radius.

As will be shown in Appendix~\ref{app.sec},
we additionally remark that the black-sphere empirical relation,
Eq.~(\ref{BSrcs.eq}), is found out to be valid within $\approx 10$\% for this antiproton case.
This will support that the same line of the discussion
in Ref.~\cite{Hatakeyama18} but for antiproton can be extended here.

\section{Conclusions}
\label{conclusion.sec}

We have investigated the feasibility
of using the antiproton-nucleus ($\bar{p}A$) 
scattering as a probe of the nuclear surface density distribution.
We have shown that the high-energy $\bar{p}A$
reactions are well described with the Glauber model
with a ``minimal'' profile function, which
reproduces the $\bar{p}$-$^{12}$C total reaction cross sections
in a wide range of the incident energies,
through a comparison to the available experimental data
of the antiproton
elastic scattering differential cross sections on
$^{12}$C, $^{16}$O, and $^{40}$Ca.

We have quantified what density regions are sensitive
to the $\bar{p}A$ scattering by comparing
the reaction probabilities obtained for the $\bar{p}A$ and $pA$
scattering. In the $pA$ scattering,
  the reaction probability becomes
  half at the radius of a sphere
  that corresponds to $\approx 1/10$ of the central density,
  whereas in the $\bar{p}A$ scattering
  the reaction probability is halved at 
  the tail region of the nuclear density distribution,
  $\approx 1/100$ of the central density.
The reaction probability beyond the nuclear half density radius
is significantly increased even at the low density
due to much larger elementary cross sections than the $NN$ ones.
This results in the large enhancement of
the total reaction cross sections, especially for light nuclei
which consist mostly by the nuclear surface.
We have shown that the enhancement
of the cross section is significant enough to determine
the density profile around the nuclear surface, the nuclear ``diffuseness''.
To explore the outer part of the density distribution of the exotic nuclei,
the sensitivity to the dilute nuclear tail has also been quantified
by taking an example of $^{31}$F, which is a candidate 
  of a two-neutron halo nucleus~\cite{Michel20,Masui20}.

The antiproton probes the dilute density distributions
around and beyond the nuclear surface more efficiently than the proton.
Measuring the both $\bar{p}A$ and $pA$
total reaction and elastic scattering cross sections
could offer the opportunity to precisely determine the nuclear
surface density profile including the dilute nuclear tail.
Experimental search for new halo candidates will extend
for heavier nuclei beyond $^{29,31}$F.
Recently, unexpectedly rapid increase of the nuclear radii of
neutron-rich calcium isotopes towards larger neutron excess
was reported~\cite{Tanaka20}.
A possible interpretation could be a drastic change
of the structure of the core nucleus and is related
to the properties of the valence single particle orbits~\cite{Horiuchi20},
which determine the nuclear diffuseness.
Though no experimental facility might exist so far doing the measurement
of the high-energy antiproton off an unstable nucleus,
if realized, as the electron scattering did~\cite{scrit2017},
the antiproton can be one of the best probes
to unveil the exotic structure of neutron-rich nuclei.

\acknowledgments

We thank H. Masui and M. Kimura for making the numerical data
of the $^{31}$F density distributions available,
and K. Iida and K. Oyamatsu for valuable communications.
This work was in part supported by JSPS KAKENHI Grants
No. 18K03635, No. 18H04569, and No. 19H05140.
We acknowledge the collaborative research program 2020, 
Information Initiative Center, Hokkaido University.

\appendix

\section{Black sphere picture in antiproton-nucleus scattering}
\label{app.sec}

The antiproton scattering offers more
absorptive scattering process than that of 
the proton scattering.
One may think that the black sphere (BS) 
model~\cite{Kohama04,Kohama05,BS3, Kohama16} 
is expected to work better than that
of the $NA$ scattering.
For this purpose, we evaluate the BS estimate
where the total reaction cross section is 
calculated by the first peak position of the diffraction peak:
Assuming that a nucleus is completely absorptive 
within a sharp-cut nuclear radius $a_{\rm BS}$,
the total reaction cross section 
\begin{align}
\sigma_{\rm BS}=\pi a_{\rm BS}^2
\label{BSrcs.eq}
\end{align}
can be related to the BS radius~\cite{Kohama04}
\begin{align}
  a_{\rm BS}=\frac{5.1356\cdots}{2p\sin(\theta_M/2)},
\label{BSsigma.fig}
\end{align}
where $p$ $(=K)$ is the momentum between the two colliding particles.
If the $\bar{N}A$ scattering is ideally described with the BS model,  
a slope of the BS cross sections must follow the $y=x$ line 
in this correlation plot.

\begin{figure}[ht]
\begin{center}
\epsfig{file=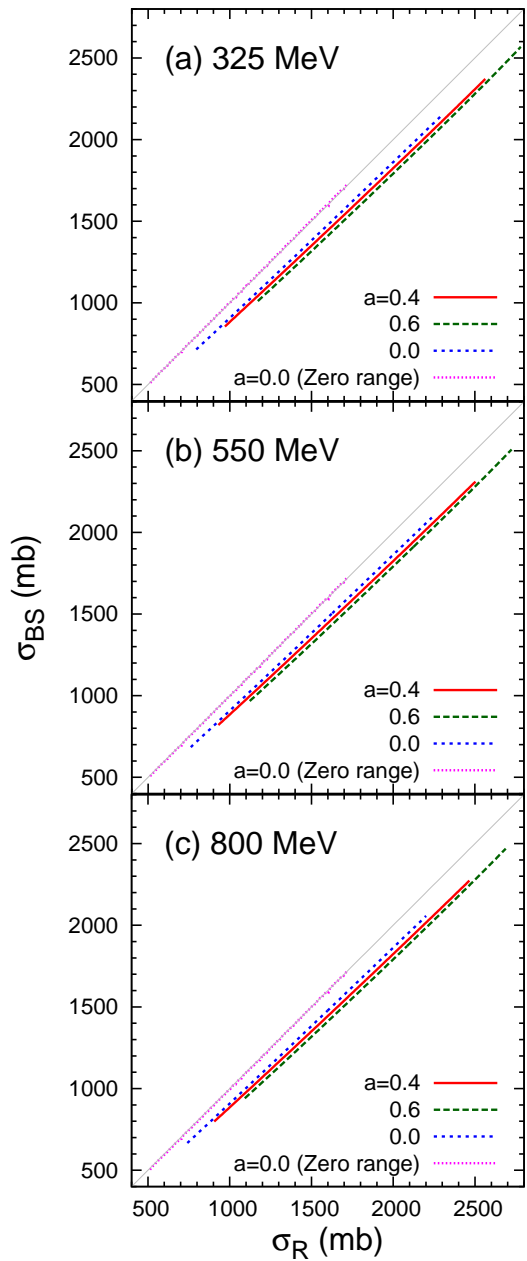, scale=1.2}
\caption{Correlation plot of $\sigma_{BS}$ versus $\sigma_R$
 at (a) 325, (b) 550, and (c) 800 MeV for the antiproton scattering.
Thin lines in gray denote a $y=x$ line for a guide of eyes.}
 \label{corrplt.fig}
\end{center}
\end{figure}

Figure~\ref{corrplt.fig} plots $\sigma_{BS}$ against $\sigma_{R}$
with the 2pF density distributions of Eq.~(\ref{Fermi.eq}).
The $\sigma_{BS}$ deviates from $\sigma_R$ 
with increasing the diffuseness parameter of the density distribution.
We note that the Glauber calculation with the zero-range profile function
is nothing but the complete absorption or the BS model 
if the elementary cross section is large enough.
Actually, as displayed in Fig.~\ref{corrplt.fig}, 
the correlation plot follows the $y=x$ line
with a sharp-cut square-well ($a=0$) using the zero-range profile function.
The deviation comes from the two facts in reality, that are,
the nuclear surface diffuseness and the finiteness of the interaction.
Though the black sphere model explains most of
the bulk properties of the $\bar{p}A$ scattering,
the differences are typically $\approx 10$\% in $A=40$--250
with $a=0.4$--0.6 fm for all the incident energies,
which are a bit larger than
the case of the $pA$ scattering~\cite{Hatakeyama18}
due to higher sensitivity to the nuclear surface.

\end{document}